\documentclass[11pt, a4paper]{article}
\usepackage[margin=2.75cm]{geometry}
\usepackage{amsmath,amsthm,amssymb, color, bm, cite}
\usepackage{graphics, graphicx}
\usepackage{epsfig,  pstricks}

\newcommand{\be}[1]{\begin{equation}#1 \end{equation}}

\newcommand{\evenodd}[1]{\left\lfloor #1 \right\rfloor}

\newcommand{\paren}[1]{\left( #1 \right)}

\newcommand{\dis}{\displaystyle}
\newcommand{\goodspace}{\hspace{0.6cm}}

\newcommand{\arcsinh}{\text{arcsinh}}

\def\eg{{\it e.g.\ }}
\def\etal{{\it et al.\ }}
\def\ie{{\it i.e.\ }}

\begin{document}



\begin{center}
\textbf{\Large On explicit thermodynamic functions and extremal limits \\[.3cm] of Myers-Perry black holes }\\
\vspace{1cm}
\normalsize \textbf{Jan E. \AA man}\footnote{ja@fysik.su.se} \normalsize{and} \normalsize {\bf Narit Pidokrajt}\footnote{narit@fysik.su.se, pidokrajt@gmail.com} \\[.2cm]
\emph{Department of Physics \\
Stockholm University\\
SE-106 91  Stockholm\\
Sweden} \\[.4cm]

\end{center}

\begin{abstract}
We study thermodynamic properties of Myers-Perry black holes by deriving explicit fundamental relations from which we can obtain the temperature and specific heat in terms of explicit control parameters in arbitrary dimensions. Using the definition of extremal black holes we establish the generalized Kerr bound in arbitrary dimension.  We study thermodynamic geometries of the Myers-Perry black holes with equal angular momenta in arbitrary dimensions and draw thermodynamic cone diagrams which capture the extremal limits of the black holes. Thermodynamic state space is represented geometrically as a wedge embedded in Minkowski space. The opening angle of such a wedge is uniquely determined by the number of spacetime dimensions and the number of angular momenta. Our results can potentially be used to generalize thermodynamic instability analysis and other studies in which extremal limits of the Myers-Perry black holes are required. 
\end{abstract}

\section{Introduction}

Black hole solutions in more than four spacetime dimensions have been the subject of increasing attention in recent years. Of particular interest are the Myers-Perry (MP) black holes~\cite{Myers:1986un} whose   uncharged rotating version is a direct generalization of the Kerr black hole solution in General Relativity.  The MP solutions are significant mainly because of the richness of the solutions themselves. This is due to   the possibility of rotation in  $N$ independent rotation planes with the rotation group $SO(d-1)$ having Cartan subgroup $U(1)^{N}$ with
\be{
N \equiv \left\lfloor
\frac{d-1}{2} \right\rfloor \, ,
}
where $\evenodd{\phantom{x}}$ denotes the integer part.  To each of these rotations there is an associated angular momentum component $J_i$.  Dimensionality also plays a role when one consider the dynamics of the black hole solutions as pointed out by Emparan and Reall \cite{Emparan:2008eg}.  The other aspect of rotation that changes qualitatively as we increase the number of dimensions is the competition between the gravitational and centrifugal potentials. The radial fall-off of the Newtonian potential 
$
-\frac{GM}{r^{d-3}}
\label{potential1}
$
depends on the number of dimensions, whereas the centrifugal barrier
$
\frac{J^2}{M^2 r^2}\,
\label{potential2}
$
does not as rotation is confined to a plane. It is readily seen that the competition between the two quantities is different in $d=4$, $d=5$, and $d\geqslant 6$. This shows that the dimensionality will have dramatic consequences for the behavior of black holes.   Vacuum black hole solutions in five dimensions include the so-called black rings when we consider stationary solutions with two rotational Killing vectors~\cite{Emparan:2001wn}. The black ring solutions in five dimensions are exact solutions having the horizon topology $S^2 \times S^1$.   The other feature that appears in higher dimensions is the presence of black objects with extended horizons, \ie  black strings and in general black $p$-branes which are unstable objects~\cite{Gregory:1993vy}. We take note that these solutions are not asymptotically flat   but they give us intuition for black holes in higher dimensions.

For a comprehensive classification of black hole species in higher dimensions we refer the reader to a review by Rodriguez~\cite{species}. Recently an interesting study of particle injections to MP black holes and black rings was done by Bouhmadi-Lopez \etal  \cite{BouhmadiLopez:2010vc} finding that this particular way of destroying a black hole is not possible and that Cosmic Censorship is preserved.

In this paper we study merely the uncharged MP black hole solutions in asymptotically flat spacetime with arbitrary rotation in each of the $N \equiv \left\lfloor \frac{d-1}{2} \right\rfloor$ independent rotation planes. The solutions have to be treated separately depending on whether the number of dimensions is odd or even.   The black holes have $(d-1)/2$ angular momenta if $d$ is odd and $(d-2)/2$ if $d$ is even. The multiple-spin Kerr black hole's metric in Boyer-Lindquist coordinates for odd $d$ is given by 
\be{
ds^2 = -d\bar{t}^2 + (r^2 + a^2_i)(d\mu^2_i + \mu^2_i d\bar{\phi}^2_i) 
+\frac{\mu r^2}{\Pi F}(d\bar{t} + a_i \mu^2_i d\bar{\phi}_i)^2 + \frac{\Pi F}{\Pi - m r^2}dr^2, 
}
where 
\be{
d\bar{t} = dt - \frac{m r^2}{\Pi - m r^2}dr,
}

\be{
d\bar{\phi}_i = d\phi_i + \frac{\Pi}{\Pi - m r^2}\frac{a_i}{r^2 + a^2_i}dr,
}
with the constraint 
\be{
F =   1 - \frac{a^2_i \mu^2_i}{r^2 + a^2_i},
}
\be{
\mu^2_i = 1.
}
The function $\Pi$ is defined as follows:
\be{
\label{eq:parameters-multi-Kerr}
\Pi = \prod_{i=1}^{(d-1)/2} (r^2 + a^2_i).
}
The metric is slightly modified for even $d$. The event horizons in the Boyer-Lindquist coordinates occur where $g^{rr} = 1/g_{rr}$ vanishes. They are the largest roots of 
\be{
\label{make1}
\Pi - m r = 0 \goodspace \text{even $d$} 
}
\be{
\label{make2}
\Pi - m r^2 = 0 \goodspace \text{odd $d$} .
}
The areas of the event horizon are given by 
\be{
A = \frac{\Omega_{(d-2)}}{r_+} \prod_i (r^2_+ + a^2_i) \goodspace \text{odd $d$},
}
\be{
A = \Omega_{(d-2)} \prod_i (r^2_+ + a^2_i) \goodspace \text{even $d$}.
}
In $d = 5$ there can be only two angular momenta associated with the Kerr black hole, thus the area of the event horizon reads
\be{
A = \frac{2\pi^2}{r_+} (r^2_+ + a_1^2)(r^2_+ + a^2_2). 
}
The Bekenstein-Hawking entropy is given by $\dis S= \frac{k_B A}{4G}$, and we can choose $k_B = {1 \over \pi}$ and $G = \frac{\Omega_{(d-2)}}{4\pi}$ so that the Bekenstein-Hawing entropy for the MP black holes is simplified as 
\be{
S  =  \frac{1}{r_+} \prod_i (r^2_+ + a^2_i) \goodspace \text{odd $d$},
}
\be{
S =  \prod_i (r^2_+ + a^2_i) \goodspace \text{even $d$}.
}

Parts of this paper are devoted to a study of extremal limits of MP black holes. Extremal black holes are those with vanishing temperature (because zero surface gravity is zero in the extremal limit).  In a thermodynamic sense such the black holes do not radiate thermally.

\section{The fundamental relation}

In this section we study and derive a fundamental relation of  MP black holes with all the allowed number of spins appearing in any number of dimensions.   We utilize the manipulation employed in~\cite{Aman:2005xk} in obtaining the fundamental relation. For the single spin case we begin by observing that 
\be{
S = r_+ m.
}
Inserting $r_+ = \frac{S}{m}$ in Eqs.~(\ref{make1}) and~(\ref{make2}) we can solve for $m$ in terms of $S$ and $a$ where 
\be{
m = \frac{4 M}{d-2},
}
and
\be{
a= \frac{d-2}{2}\frac{J}{M}.
} 
Next we can solve for $M$ which is given by 
\be{
\label{eq:Kerr-Mass}
M = \frac{d-2}{4}S^{\frac{d-3}{d-2}} \paren{1 + \frac{4J^2}{S^2}}^{1/(d-2)}.
}
In the multiple-spin case we can follow the same procedure used in the single-spin case but we now have
\be{
a_i= \frac{d-2}{2}\frac{J_i}{M}\,,
}
and the desired mass formula $M(S,J_1, J_2, ..., J_n)$ is found to be
\begin{equation} 
M = \frac{d-2}{4} S^\frac{d-3}{d-2}  \prod_i^n \paren{ 1+\frac{4J_i^2}{S^2} }^\frac{n}{d-2}.
\end{equation}
Zero spins can be ignored but the number $n$ must obey $n \leqslant \evenodd{\frac{d-1}{2}}$. Performing  differentiation, $T =\frac{\partial M}{\partial S}$, we obtain the black hole's temperature as
\be{
\label{temperature}
\dis
T =  \frac {\dis \frac{d-3}{4} \prod_i^n \paren{1+\frac{4J_i^2}{S^2} }
          -\frac{1}{2} \sum_k^n \frac{4J_k^2}{S^2} \dis
                      \prod_{i \ne k}^n \paren{ 1+\frac{4J_i^2}{S^2} } }
        {\dis S^\frac{1}{d-2} \prod_i^n \paren{ 1+\frac{4J_i^2}{S^2}}^\frac{d-3}{d-2}}.
}
As an example we will compute the mass function and the black hole temperature in $d=7$ with three independent angular momenta
\begin{equation} 
M = \frac{5}{4} S^\frac{4}{5}
      \paren{ 1 +\frac{4J_1^2}{S^2}}^\frac{1}{5}
      \paren{ 1 +\frac{4J_2^2}{S^2}}^\frac{1}{5}
      \paren{ 1 +\frac{4J_3^2}{S^2}}^\frac{1}{5}.
\end{equation}
The temperature in $d=7$ is as follows:
\begin{equation} 
\label{threespins}
T =  \frac{\partial M}{\partial S} =
  \frac {2 +\frac{4J_1^2}{S^2} +\frac{4J_2^2}{S^2} +\frac{4J_3^3}{S^2}
           -\frac{4J_1^2}{S^2} \frac{4J_2^2}{S^2} \frac{4J_3^2}{S^2}}
        {2 S^\frac{1}{5} \paren{ 1 +\frac{4J_1^2}{S^2}}^\frac{4}{5}
                         \paren{ 1 +\frac{4J_2^2}{S^2}}^\frac{4}{5}
                         \paren{ 1 +\frac{4J_3^2}{S^2}}^\frac{4}{5}},
\end{equation}
which becomes zero in the extremal limit.

\begin{figure}[ht]
\begin{minipage}[b]{0.5\linewidth}
\centering
\includegraphics[scale=.4]{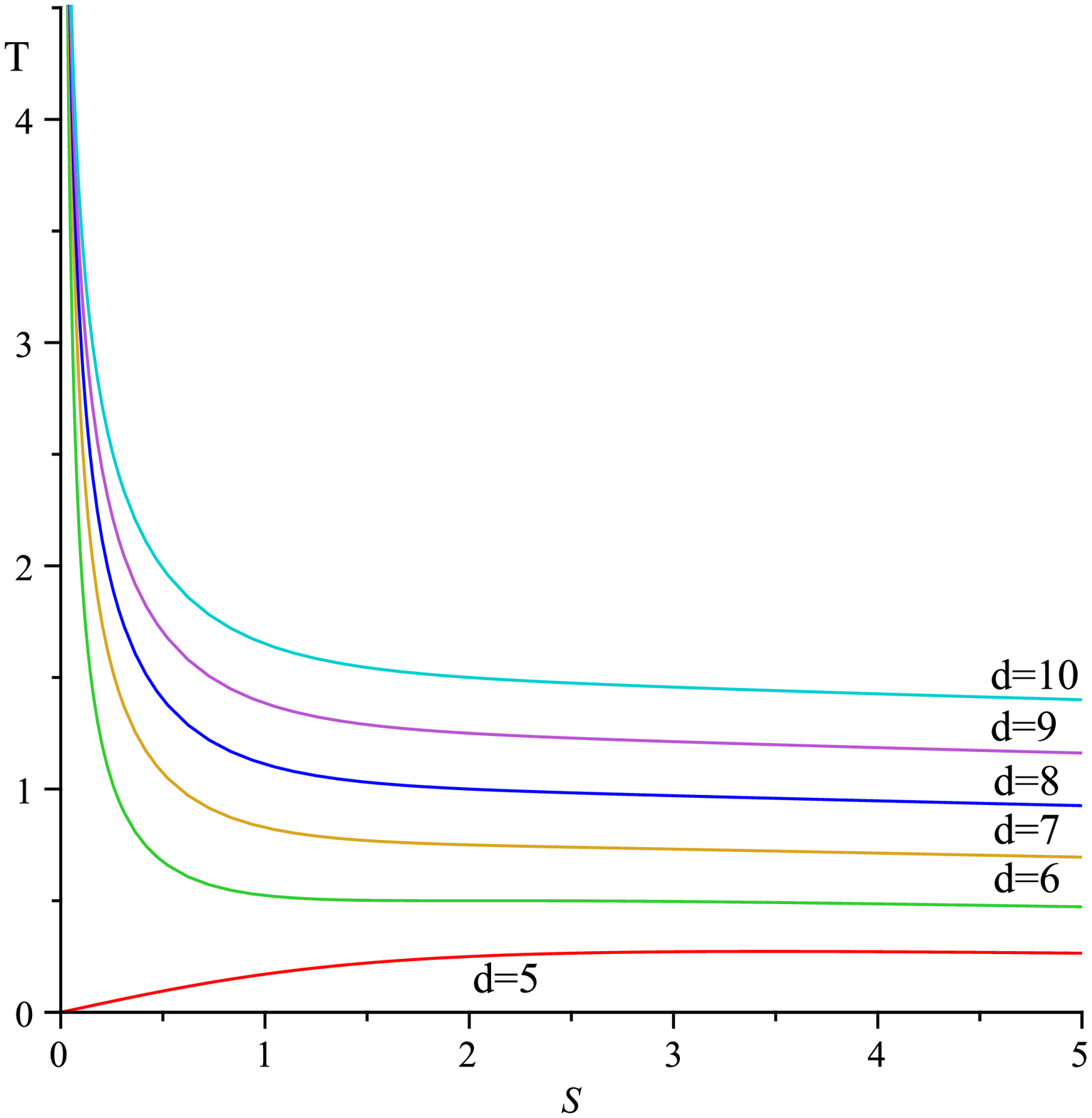}
\caption{\sl \small Temperatures of the MP black holes (at fixed $J$) in various dimensions when there is only one angular momentum present. Note that in $d=5$ the temperature vanishes at zero entropy, whereas in other dimensions  the temperatures tend to infinity showing that there are no extremal limits for the black hole solutions in $d>5$ when there is only one angular momentum turned on. }
\label{fig:single_j}
\end{minipage}
\hspace{0.4cm}
\begin{minipage}[b]{0.5\linewidth}
\centering
\includegraphics[scale=.4]{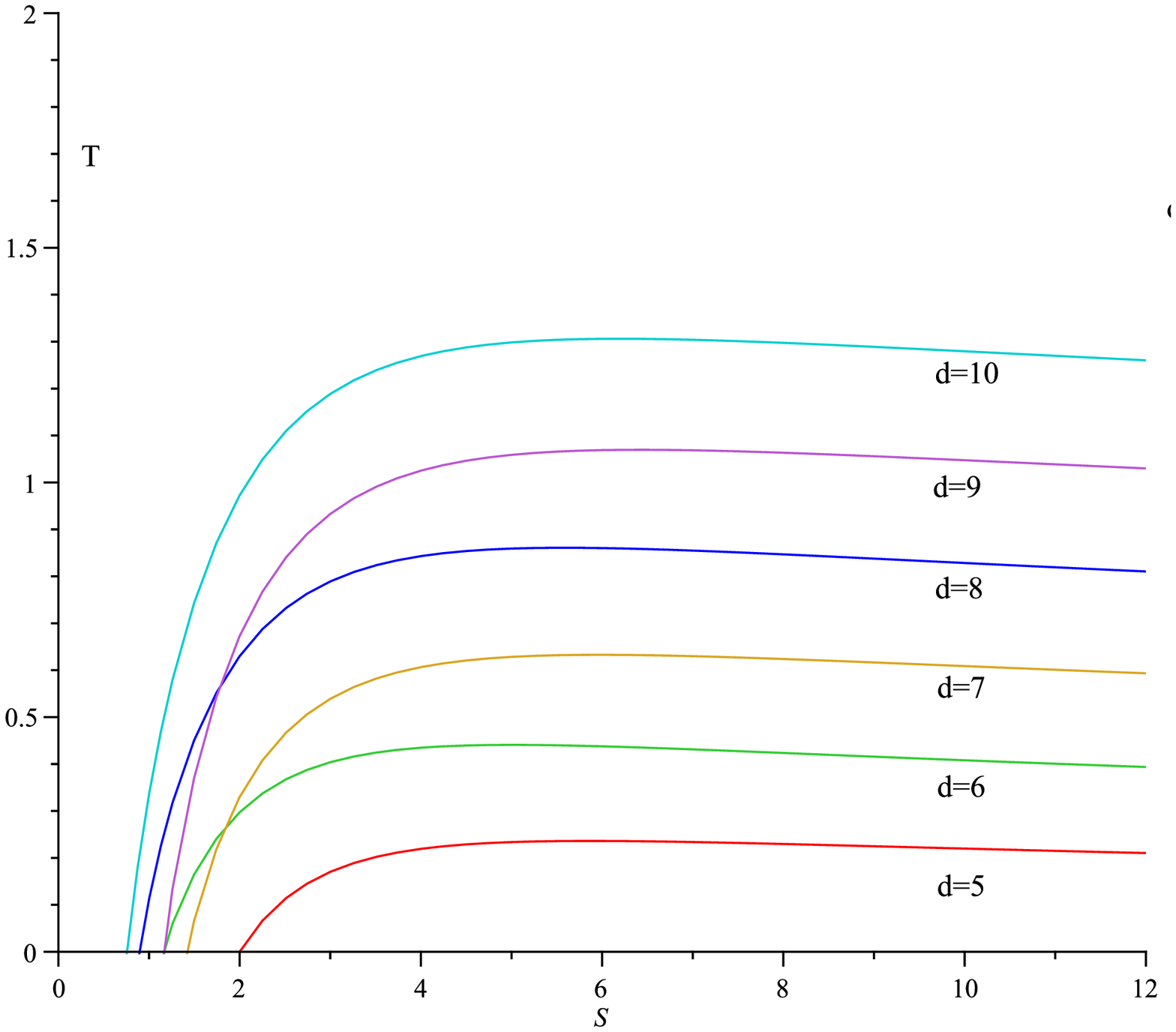}
\caption{\sl \small   Temperatures of the MP black holes  (at fixed $J$) in various dimensions when all the angular momenta are present and they are equal in magnitude. The extremal limits can be read off at $T=0$. As the number of dimension increases the extremal limits tend to occur at lower entropies (but note \eg the irregularities of how $T$ approaches zero for $T=0$ between $d=6, 7$ and $d=8,9$. }
\label{fig:equal_j}
\end{minipage}
\end{figure}

We  study thermodynamic functions of the MP black holes and for simplicity concentrate on two special cases \ie 
when (i) when a certain number of spins are zero and the remaining spins are nonzero and equal (ii) when all spins are turned on and are equal in magnitude.  We will make attempts to investigate more general cases in forthcoming papers. 

\subsection{All nonzero spins $J$ equal}

For the particular case when we have $n$ nonzero equal spins $J$, and remaining spins are zero we can write the mass function as 

\begin{equation} 
M = \frac{d-2}{4} S^\frac{d-3}{d-2}\paren{ 1+\frac{4J^2}{S^2}}^\frac{n}{d-2}
\end{equation}
where $n \leqslant \lfloor\frac{d-1}{2}\rfloor$.  The temperature becomes after factorization
\begin{equation} 
T = \frac{\partial M}{\partial S} =   \frac{(d-3) \paren{ 1 +2\sqrt\frac{2n-d+3}{d-3}\frac{J}{S}} \paren{1 -2\sqrt\frac{2n-d+3}{d-3}\frac{J}{S}}}
       {4 S^\frac{1}{d-2} \paren{1 +\frac{4J^2}{S^2}}^\frac{d-n-2}{d-2}}\,.
\end{equation}
When $T = 0$ we have an extremal limit.  If $2n \geqslant d - 3$ there will be an extremal limit at
\begin{equation} 
\frac{S}{J}\Big|_{\small extr} =  2 \sqrt\frac{2n-d+3}{d-3}\,.
\end{equation}
This limit exists for $2n > d - 3$ which can be expressed in $(M, S)$ coordinates as 
\begin{equation}
\frac{S^{d-3}} {M^{d-2}}\Big|_{\small extr} =  \frac{2^{2d-n-4} (2n-d+3)^n} {(d-2)^{d-2}n^n}
\end{equation}
or in $(M, J)$ coordinates 
\begin{equation}
\frac{J^{d-3}} {M^{d-2}}\Big|_{\small extr}  =  \frac{2^{d-n-1} (2n-d+3)^\frac{2n-d+3}{2}(d-3)^\frac{d-3}{2}} {(d-2)^{d-2}n^n}\,.
\end{equation}
For the case $2n = d -3$ case (requiring odd $d$)  we have $S_{\small ext} = 0$ and 
\be{
\frac{J^{d-3}} {M^{d-2}}\Big|_{\small extr} = \frac{2^{d-1}} {(d-2)^{d-2}}\,.
}
For even $d$ with all $n = \frac{d}{2} -1$ spins we can also express 
\be{
\frac{S} { J}\Big|_{\small extr} = \frac{2}{\sqrt{d-3}},
}
and
\be{
\frac{J^{d-3}} {M^{d-2}}\Big|_{\small extr} =  \frac{2^{d-1} (d-3)^\frac{d-3}{2}} {(d-2)^\frac{3(d-2)}{2}}\,.
}
For odd $d$ with all $n = \frac{d-1}{2}$ spins we have 
\be{
\frac{S} { J}\Big|_{\small extr} = \frac{2\sqrt{2}}{\sqrt{d-3}},
}
and
\be{
\frac{J^{d-3}} {M^{d-2}}\Big|_{\small extr} = \frac{2^d (d-3)^\frac{d-3}{2}} {(d-2)^{d-2} (d-1)^\frac{d-1}{2}}.
}
The Schwarzschild limit is when $J = 0$ and this sets a physical bound to be 
\be{
\frac{S^{d-3}} {M^{d-2}} \leqslant \frac{4^{d-2}} {(d-2)^{d-2}}.
}
We list the $\frac{S}{J}\big|_{\small extr}$ for $d \leqslant 11$ in Table~\ref{table1}. We present extremal limits in various coordinates for $d$ up to 11 in Appendix B.  

\begin{table}
\begin{center}
\begin{tabular}{|r|r@{ }l|r@{ }l|r@{ }l|r@{ }l|r@{ }l|}
\hline 
$d$ & $n$ & $=1$ & $n$ & $=2$ & $n$ & $=3$ & $n$ & $=4$ & $n$ & $=5$   \\
\hline 
4  & $\frac{S}{J}$ & $\geqslant 2$                         & & & & & & & & \\[.1cm]
5  & $\frac{S}{J}$ & $\geqslant 0$ & $\frac{S}{J}$ & $\geqslant 2$   & & & & & & \\[.1cm]
6  & - & & $\frac{S}{J}$ & $\geqslant \frac{2}{\sqrt{3}}$        & & & & & & \\[.1cm]
7  & -& & $\frac{S}{J}$ & $\geqslant 0$ & $\frac{S}{J}$ & $\geqslant \sqrt{2}$ & & & & \\[.1cm]
8  & -& &- & & $\frac{S}{J}$ & $\geqslant \frac{2}{\sqrt{5}}$        & & & & \\[.1cm]
9  & -& &- & & $\frac{S}{J}$ & $\geqslant 0$ & $\frac{S}{J}$ &
                             $\geqslant \frac{2}{\sqrt{3}}$            & & \\[.1cm]
10 &- & &- & &- & & $\frac{S}{J}$ & $\geqslant \frac{2}{\sqrt{7}}$        & & \\[.1cm]
11 & -& & -& & -& & $\frac{S}{J}$ & $\geqslant 0$ & $\frac{S}{J}$ & $\geqslant 1$   \\[.1cm]
\hline
\end{tabular}
\end{center}
\caption{\sl Extremal limits in $(S, J)$ coordinates for MP black holes in various spectrum of dimensions, $d$, for different angular momenta, $n$. A dash means that no extremal limit exists.  }
\label{table1}
\end{table}

\subsubsection{Specific heat}

It is straightforward to compute the specific heat for MP black holes of dimension $d$ with $n$ equal spins is
\begin{equation}
C = \frac{T}{\frac{\partial^2 M}{\partial S^2}} =
    \frac{(d-2)(1+\frac{4J^2}{S^2})(1-\frac{2n-d+3}{d-3}\frac{4J^2}{S^2})S}
         {\frac{(2n-d+3)(2n+1)}{d-3}(\frac{4J^2}{S^2})^2
          +\frac{2(dn-d+3)}{d-3}\frac{4J^2}{S^2} -1}
\end{equation}
It goes to infinity at
\begin{equation}
\frac{4J^2}{S^2} =
  \frac{dn-d+3-\sqrt{n(d^2n+4dn-4d^2-12n+20d-24)}}{2dn+d-4n^2-8n-3},
\end{equation}
and except for $2n = d-3$ it has a zero at
\begin{equation}
\frac{4J^2}{S^2} = \frac{d-3}{2n-d+3} .
\end{equation}

\begin{figure}[ht]
\centering
\includegraphics[scale=.8]{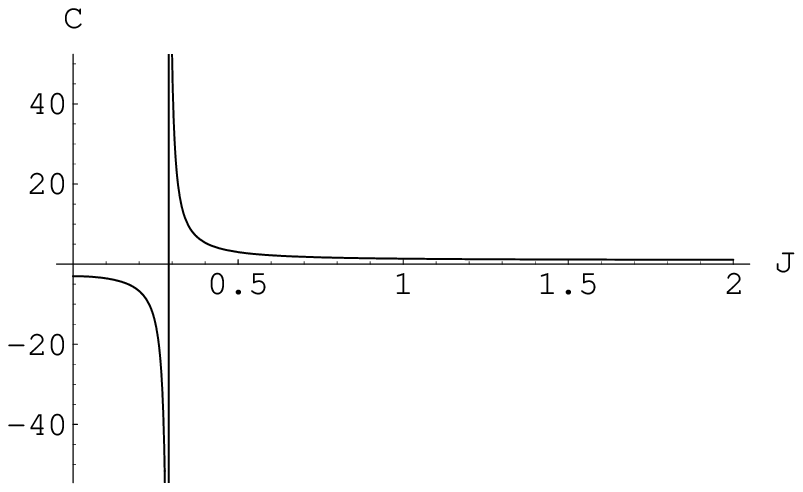}
\caption{\small \sl A plot of $C$ as a function of $J$ for $d=5$ , $n=1$ with $S=1$. It goes to infinity at $J =\frac{1}{2\sqrt{3}}$. A plot of $C$ as a function of $J$ for $d=5$ , $n=2$ with $S=1$.}
\includegraphics[scale=.8]{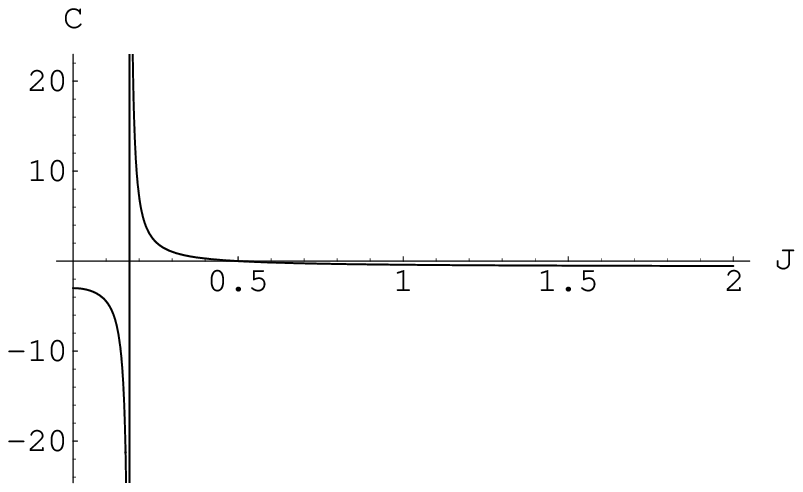}
\caption{\small \sl  A plot of $C$ as a function of $J$ for $d=5$ , $n=2$ with $S=1$. The specific heat in this case changes sign at $ J =\frac{1}{2}$ and goes to infinity at $ J =\sqrt\frac{\sqrt{21}-4}{20}$.}
\end{figure}

\subsection{All spins $J$ equal}

For the particular case when all allowed $\evenodd{\frac{d-1}{2}}$ spins $J$ are equal we still have
\begin{equation} 
M = \frac{d-2}{4} S^\frac{d-3}{d-2} \paren{ 1+\frac{4J^2}{S^2} }^\frac{n}{d-2},
\end{equation}
where $n = \left \lfloor \frac{d-1}{2} \right \rfloor$ . For even $d$ we have $n = \frac{d}{2} -1$ and the temperature after factorization becomes
\begin{equation} 
T =  \frac{(d-3) \paren{1 +\frac{2J}{\sqrt{d-3}S}} \paren{1 -\frac{2J}{\sqrt{d-3}S}}}{4 S^\frac{1}{d-2} \paren{1+\frac{4J^2}{S^2}}^\frac{1}{2}},
\end{equation}
while for odd $d$ we have $n = \frac{d-1}{2}$ and temperature
\begin{equation} 
T =   \frac{(d-3) \paren{ 1 +\frac{2\sqrt{2}J}{\sqrt{d-3}S}} \paren{1 -\frac{2\sqrt{2}J}{\sqrt{d-3}S}}}  {4 S^\frac{1}{d-2} \paren{1+\frac{4J^2}{S^2}}^\frac{d-3}{2(d-2)}}\,.
\end{equation}

\section{Thermodynamic geometries}

In this section we study thermodynamic geometry (also known as Ruppeiner geometry~\cite{Ruppeiner:1995zz}) of the MP black hole families. Black hole thermodynamic geometry has been studied over the past decade, see \eg \cite{ruppeinergeometry} and references therein for a review.  The geometrical patterns are given by the curvature of the Ruppeiner metric defined as the Hessian of the entropy on the state space of the thermodynamic system 
\begin{equation}
g^R_{ij} = - \partial_i \partial_j S(M, N^a),
\end{equation}
where $M$ denotes mass (internal energy) and $N^a$ are other mechanically conserved parameters such as charge and spin. The indices $i,j$ run over these parameters. The minus sign ensures that the metric has positive signature when the entropy function is concave.  This metric is conformal to the so-called Weinhold  metric (defined as the Hessian of energy function) via $g^W_{ij} = T g^R_{ij}$ where $T$ is thermodynamic temperature of the system of interest.  In ordinary thermodynamics it has been argued that the curvature scalar of the Ruppeiner metric measures the complexity of the underlying interactions of the system, \ie the metric is flat for the ideal gas whereas it has curvature singularities for the van der Waals gas\footnote{In this case the singularities are associated with phase transitions}. However the story is different in black hole thermodynamics in that results have been obtained but there has not been a consensus on how to interpret uncovered geometrical patterns of black hole thermodynamics.  In~\cite{arcioni} it is argued that local thermodynamic instability of black holes is encoded in the Ruppeiner metric, and that this method is consistent with the Poincar\'e method of stability analysis.  Other works in this direction can be found in~\cite{rupp0, rupp1, rupp2, rupp3, rupp4, rupp6, rupp7, rupp9, rupp10, rupp11, rupp12, rupp13, rupp14, rupp15, rupp16, rupp17, rupp18, rupp19, rupp20, rupp21, rupp22}. 

Apart from being a tool to analyze black hole's stability, it is expected that geometrical patterns (whether the metric is flat or nonflat) will play a role in the context of quantum gravity. Below we study both Ruppeiner and Weinhold geometries of the MP black holes in arbitrary dimensions. Instability of MP black holes have been investigated \eg in~\cite{stab1, stab2, stab3}.   Recently in~\cite{maria} Astefanessei, Rodriguez and Theisen argue that the singularity of the Ruppeiner metric could help detect the threshold of the membrane phase of MP black holes. In particular they study the Ruppeiner curvature of doubly spinning MP black holes in arbitrary dimensions. Their results will be discussed in comparison with ours in this section.

We use an algebraic computation package CLASSI~\cite{classi} in computing all the metrics and associated curvature scalars.

\subsection{Weinhold and Ruppeiner metrics}

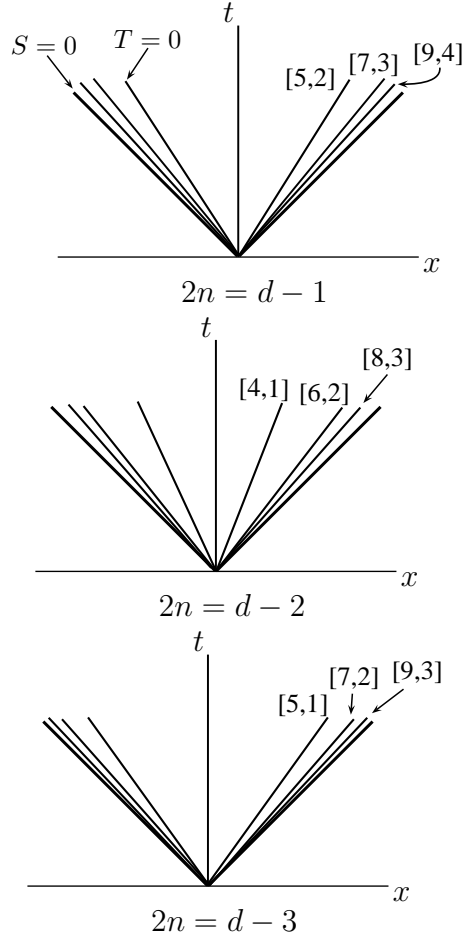
\begin{figure}[ht]
\centering
\scalebox{1} 
{
\begin{pspicture}(0,-2.0746875)(6.1078124,2.0746875)
\psline[linewidth=0.018cm](0.6821875,-1.40125)(5.4821873,-1.40125)
\psline[linewidth=0.025999999cm](3.0821874,-1.40125)(3.0821874,1.67875)
\psline[linewidth=0.04cm](3.0932431,-1.3901944)(5.271132,0.78769445)
\psline[linewidth=0.04cm](3.071132,-1.3901944)(0.8932431,0.78769445)
\usefont{T1}{ptm}{m}{n}
\rput(2.9595313,1.85375){\large $t$}
\usefont{T1}{ptm}{m}{n}
\rput(5.659531,-1.50625){\large $x$}
\usefont{T1}{ptm}{m}{n}
\rput(3.2995312,-1.86625){\large $2n = d - 1$}
\usefont{T1}{ptm}{m}{n}
\rput{-0.506288}(-0.008148629,0.035990693){\rput(4.041094,0.95875){\small [5,2]}}
\usefont{T1}{ptm}{m}{n}
\rput{-0.27862406}(-0.00548641,0.023884945){\rput(4.881094,1.15875){\small [7,3]}}
\usefont{T1}{ptm}{m}{n}
\rput(5.7410936,1.29875){\small [9,4]}
\psline[linewidth=0.027999999cm](4.5621877,0.95875)(3.0934045,-1.3833224)
\psline[linewidth=0.027999999cm](1.5821875,0.93875)(3.0909705,-1.3833224)
\psline[linewidth=0.025999999cm](5.026978,0.97330505)(3.097397,-1.375805)
\psline[linewidth=0.025999999cm](1.1573972,0.97330505)(3.086978,-1.375805)
\psline[linewidth=0.027999999cm](5.1621876,0.89875)(3.0854046,-1.3816527)
\psline[linewidth=0.027999999cm](0.9821875,0.91875)(3.0789704,-1.3816527)
\psbezier[linewidth=0.022,arrowsize=0.05291667cm 2.0,arrowlength=1.4,arrowinset=0.4]{<-}(5.2021875,0.89875)(5.3421874,0.89875)(5.7821875,0.89576495)(5.7621875,1.11875)
\usefont{T1}{ptm}{m}{n}
\rput(1.8796875,1.49875){\small $T = 0$}
\psline[linewidth=0.02cm,arrowsize=0.05291667cm 2.0,arrowlength=1.4,arrowinset=0.4]{<-}(1.6021875,0.95875)(1.8821875,1.37875)
\usefont{T1}{ptm}{m}{n}
\rput(0.4996875,1.41875){\small $S = 0$}
\psline[linewidth=0.02cm,arrowsize=0.05291667cm 2.0,arrowlength=1.4,arrowinset=0.4]{<-}(0.8821875,0.83875)(0.5421875,1.27875)
\end{pspicture} 
}
\scalebox{1} 
{
\begin{pspicture}(0,-2.0746875)(5.3378124,2.0746875)
\psline[linewidth=0.018cm](0.0,-1.40125)(4.8,-1.40125)
\psline[linewidth=0.025999999cm](2.4,-1.40125)(2.4,1.67875)
\psline[linewidth=0.04cm](2.4110556,-1.3901944)(4.5889444,0.78769445)
\psline[linewidth=0.04cm](2.3889444,-1.3901944)(0.21105556,0.78769445)
\usefont{T1}{ptm}{m}{n}
\rput(2.2773438,1.85375){\large $t$}
\usefont{T1}{ptm}{m}{n}
\rput(4.9773436,-1.50625){\large $x$}
\usefont{T1}{ptm}{m}{n}
\rput(2.6173437,-1.86625){\large $2n = d - 2$}
\usefont{T1}{ptm}{m}{n}
\rput{-0.506288}(-0.008717932,0.027137427){\rput(3.0389063,1.01875){\small [4,1]}}
\usefont{T1}{ptm}{m}{n}
\rput{-0.27862406}(-0.004526155,0.018814538){\rput(3.8389063,0.95875){\small [6,2]}}
\usefont{T1}{ptm}{m}{n}
\rput(4.6789064,1.41875){\small [8,3]}
\psline[linewidth=0.027999999cm](3.28,0.83875)(2.402148,-1.406308)
\psline[linewidth=0.027999999cm](1.36,0.85875)(2.397852,-1.3863081)
\psline[linewidth=0.027999999cm](4.08,0.77875)(2.4070554,-1.3940102)
\psline[linewidth=0.027999999cm](0.64,0.79875)(2.3929446,-1.3940102)
\psline[linewidth=0.027999999cm](4.32,0.77875)(2.40607,-1.3846836)
\psline[linewidth=0.027999999cm](0.44,0.81875)(2.3368046,-1.3539038)
\psline[linewidth=0.02cm,arrowsize=0.05291667cm 2.0,arrowlength=1.4,arrowinset=0.4]{<-}(4.36,0.83875)(4.68,1.21875)
\end{pspicture} 
}
\scalebox{1} 
{
\begin{pspicture}(0,-2.0746875)(5.545625,2.0746875)
\psline[linewidth=0.018cm](0.0,-1.40125)(4.8,-1.40125)
\psline[linewidth=0.025999999cm](2.4,-1.40125)(2.4,1.67875)
\psline[linewidth=0.04cm](2.4110556,-1.3901944)(4.5889444,0.78769445)
\psline[linewidth=0.04cm](2.3889444,-1.3901944)(0.21105556,0.78769445)
\usefont{T1}{ptm}{m}{n}
\rput(2.2773438,1.85375){\large $t$}
\usefont{T1}{ptm}{m}{n}
\rput(4.9773436,-1.50625){\large $x$}
\usefont{T1}{ptm}{m}{n}
\rput(2.6173437,-1.86625){\large $2n = d - 3$}
\usefont{T1}{ptm}{m}{n}
\rput{-0.506288}(-0.008517001,0.032615136){\rput(3.6589062,0.99875){\small [5,1]}}
\usefont{T1}{ptm}{m}{n}
\rput{-0.27862406}(-0.0064653982,0.021250712){\rput(4.3389063,1.35875){\small [7,2]}}
\usefont{T1}{ptm}{m}{n}
\rput(5.1789064,1.43875){\small [9,3]}
\psline[linewidth=0.027999999cm](3.9941943,0.8427679)(2.4058058,-1.3652679)
\psline[linewidth=0.027999999cm](0.8058058,0.8427679)(2.3941941,-1.3652679)
\psline[linewidth=0.027999999cm](4.34,0.81875)(2.418559,-1.3937509)
\psline[linewidth=0.027999999cm](0.46,0.81875)(2.421441,-1.3937509)
\psline[linewidth=0.027999999cm](4.5175285,0.8382289)(2.4024713,-1.4007288)
\psline[linewidth=0.027999999cm](0.28247124,0.8382289)(2.3975286,-1.4007288)
\psline[linewidth=0.02cm,arrowsize=0.05291667cm 2.0,arrowlength=1.4,arrowinset=0.4]{<-}(4.3,0.87875)(4.32,1.15875)
\psline[linewidth=0.02cm,arrowsize=0.05291667cm 2.0,arrowlength=1.4,arrowinset=0.4]{<-}(4.58,0.89875)(5.0,1.21875)
\end{pspicture} 
}
\caption{\sl We present the black hole solutions in three series: Series 1 is when $2n = d-1$, Series 2 when $2n = d-2$, and Series 3 when $2n = d-3$. For each series we have three cases as shown in the figures. Numbers in the brackets refer to [$d, n$] with $d$ the spacetime dimension, and $n$ the number of nonzero spins.  }
\label{figures}
\end{figure}

The Weinhold metric in original coordinates is 
\begin{eqnarray}
\label{weinhold1}
\hspace{-15mm} ds_W^2 &=&\lambda_W \Big ([-16(2n+1)(d-2n-3)J^4
                                    +8(dn-d+3)J^2S^2-(d-3)S^4]dS^2\nonumber \\
            && \hspace{+10mm} +  [64n(d-2n-3)J^3S -16n(d-1)JS^3]dSdJ\nonumber \\
            &&  \hspace{+20mm} +[-32n(d-2n-2)J^2S^2 +8n(d-2)S^4]dJ^2 \Big)
\end{eqnarray}
where
\begin{equation}
\lambda_W = \frac{1}{4(d-2)(S^2+4J^2)^\frac{2d-n-4}{d-2} S^\frac{d+2n-1}{d-2}}\,.
\end{equation}
For $n=1$ this reduces to (37)-(38) in~\cite{Aman:2005xk}. With the  coordinate transformations 
\be{
u = \frac{J}{S}
}
\begin{equation}
\tau = \sqrt\frac{d-2}{d-3} S^\frac{d-3}{2(d-2)} (1+4u^2)^\frac{n}{2(d-2)},
\end{equation}
it becomes diagonal and reads
\begin{equation}
ds_W^2 = -d\tau^2 +\frac{2n \paren{ d -3 -(d-3-2n)4u^2}}{(d -2)(1+4u^2)^2}\tau^2 du^2.
\end{equation}
This is a flat metric. It can be brought to Rindler coordinates
\begin{equation}
ds_W^2 = -d\tau^2 +\tau^2 d\sigma^2
\end{equation}
by an additional coordinate transformation 
\begin{equation}
\dis \sigma =   \int_0^u \sqrt\frac{2 n[d -3 +(2n-d+3)4u^2]}{(d -2)(1+4u^2)^2} du.
\end{equation}
Finally we can transform the metric into Minkowski coordinates $ds_W^2 = -dt^2 +dx^2$ using 
\begin{equation}
t = \tau \cosh\sigma, \goodspace x = \tau \sinh\sigma \,.
\end{equation} 
The transformation from $u$ to $\sigma$ is best studied in three subcases depending on $2n$ is greater than, equal or less than $d-3$. 
The flat metrics can be embedded as a {\it state space wedge} in the Minkowskian-like diagram, which we call {\it thermodynamic cone}\footnote{This resembles relativity's light cone structure in the sense that the cone displays the causality of the thermodynamic state space. }. The black hole temperature vanishes on the edge of the wedge, whilst the entropy vanishes on the thermodynamic cone. Hence this diagram can be used to decide which black hole families possess genuine extremal limits, \ie the black hole families without extremal limits will have no state space wedges in the thermodynamic cone, see Fig.~\ref{figures}.

\subsection*{The case $2n>d-3$}

Here the transformation is

\be{
\begin{split}
\sigma &= \frac{\sqrt{2n-d+3}}{\sqrt{2}\sqrt{d-2}}          \arcsinh\frac{2\sqrt{2n-d+3}\ u}{\sqrt{d-3}}  \\
         & \hspace*{+30mm} + \frac{\sqrt{d-n-3}}{\sqrt{d-2}}              \arctan\frac{2\sqrt{d-n-3}\ u}{\sqrt{d-3+4(2n-d+3)u^2}}.
\end{split}
}
\noindent As here $u_{\small extr} = \frac{\sqrt{d-3}}{2\sqrt{2n-d+3}}$ we have
\begin{equation}
\sigma_{\small extr} = \frac{\sqrt{2n-d+3}}{\sqrt{2}\sqrt{d-2}} \arcsinh 1
               +\frac{\sqrt{d-n-3}}{\sqrt{d-2}}
                  \arctan\frac{\sqrt{2}\sqrt{d-n-3}\ u}{\sqrt{d-3}}
\end{equation}
See also the appendix for further specifications on $2n=d-2$ and $2n=d-1$.

\begin{table}[h]
\begin{center}
\begin{tabular}{|r|r|l@{}l|l|}
\hline
$d$ & $n$ & $\sigma_{extr}$ & & $|\frac{x}{t}| \leqslant$ \\
\hline
4  & 1 & $\frac{1}{2} \arcsinh 1$ & $\approx 0.4407$ & 0.4142 \\
5  & 1 & $\frac{\pi}{2 \sqrt{3}}$ & $\approx 0.9069$ & 0.7196 \\
5  & 2 & $\sqrt\frac{2}{3} \arcsinh 1$ & $\approx 0.7196$ & 0.6167 \\
6  & 2 & $\frac{1}{2} (\arcsinh 1 +\frac{\pi}{2 \sqrt{2}})$ &
         $\approx 0.9960$ & 0.7599 \\
7  & 2 & $\frac{\pi}{\sqrt{5}}$ & $\approx 1.4050$ & 0.8864 \\
7  & 3 & $\sqrt\frac{3}{5} (\arcsinh 1 +\arctan\frac{1}{\sqrt{2}})$ &
         $\approx 1.1595$ & 0.8209 \\
8  & 3 & $\frac{1}{2} (\arcsinh 1 +2 \arctan\sqrt{2})$ &
         $\approx 1.3960$ & 0.8845 \\
9  & 3 & $\frac{3}{2 \sqrt{7}} \pi$ & $\approx 1.7811$ & 0.9448 \\
9  & 4 & $\frac{2}{\sqrt{7}} (\arcsinh 1 +\frac{\pi}{2\sqrt{2}})$ &
         $\approx 1.5059$ & 0.9062 \\
10 & 4 & $\frac{1}{2} (\arcsinh 1 +\sqrt{\frac{2}{3}} \pi)$ &
         $\approx 1.7232$ & 0.9383 \\
11 & 4 & $\frac{2 \pi}{3}$ & $\approx 2.0944$ & 0.9701 \\
11 & 5 & $\frac{\sqrt{5}}{3}(\arcsinh 1
                            +\sqrt{3} \arctan\sqrt\frac{3}{2})$ \ \ &
         $\approx 1.8009$ & 0.9469 \\
12 & 5 & $\frac{1}{2} (\arcsinh 1 +2 \sqrt{2} \arctan 2)$ &
         $\approx 2.0064$ & 0.9645 \\
\hline
\end{tabular}
\end{center}
\caption{A table showing the opening angles of a thermodynamic wedge embedded in a thermodynamic cone for $2n \geqslant d-3$. Note that as the number of angular momenta increases the wedge tends to fill up the thermodynamic cone.}
\label{table2}
\end{table}

\subsection*{The case $2n=d-3$}

This applies to dimension 5 with 1 spin, dimension 7 with 2 spins etc. We use the following transformation
\begin{equation}
\sigma = \frac{\sqrt{d-3}}{\sqrt{2}\sqrt{d-2}}\ \arctan\ 2u
\end{equation}
\begin{equation}
u = \frac{1}{2} \tan\left(\frac{\sqrt{2}\sqrt{d-2}}{\sqrt{d-3}}\ \sigma\right)
\end{equation}
Here we obtain $u_{\small extr} = \infty$ and we have
\begin{equation}
\sigma_{\small extr} = \frac{\pi\sqrt{d-3}}{2\sqrt{2}\sqrt{d-2}}
\end{equation}

\subsection*{The case $2n < d -3$}

Here there are no extremal limits as $T$ cannot become 0. Integration gives $\sigma$ as
\be{
\begin{split}
\sigma &= -\frac{\sqrt{d-2n-3}}{\sqrt{2}\sqrt{d-2}}              \arcsin\frac{2\sqrt{d-2n-3}\ u}{\sqrt{d-3}} \\
        & \hspace*{+30mm}+\frac{\sqrt{d-n-3}}{\sqrt{d-3}}        \arctan\frac{\sqrt{d-n-3}\ u}{\sqrt{d-3-4(d-2n-3)u^2}}
\end{split}
}
This formula will however not hold up to $u \rightarrow \infty$. \\

We present additional special cases in Appendix A.

\subsection{Ruppeiner metric}

Given the conformal transformation, $ds_R^2 = \frac{1}{T}ds_W^2$, the Ruppeiner metric is easily obtained in the same coordinates ($S$, $J$) and differs from the Weinhold metric in Eq.~(\ref{weinhold1}) with $\lambda_W$ replaced by
\begin{equation}
\lambda_R = \frac{1}{(d-2)(d-3)S^5(1+\frac{4J^2}{S^2})^\frac{d-4}{d-2}
                   (1 +\frac{2n-d+3}{d-3}\frac{4J^2}{S^2})}\,.
\end{equation}
The Ruppeiner metric is not flat and its curvature is given by
\begin{equation}
R_R = -\frac{1}{S}
       \frac{1 +3\frac{2n-d+3}{d-3}\frac{4J^2}{S^2}}
            {(1 +\frac{2n-d+3}{d-3}\frac{4J^2}{S^2})
             (1 -\frac{2n-d+3}{d-3}\frac{4J^2}{S^2})}\,.
\end{equation}
If $2n > d -3$ the curvature diverges at the extremal limit
\begin{equation}
\frac{J}{S} = \frac{\sqrt{d-3}}{2\sqrt{2n-d+3}},
\end{equation}
whereas if $2n < d -3$ the curvature will diverge at
\begin{equation}
\frac{J}{S} = \frac{\sqrt{d-3}} {2\sqrt{d-2n-3} }
\end{equation}
which is however not an extremal limit.   For the $2n=d-3$ case the curvature is reduced to $R_R = -\frac{1}{S}$ without a divergence.

\section{Discussions}

In this paper it is shown that MP black holes of dimension $d$ with $n$ equal nonzero spins and $2n \geqslant d-3$ all have extremal limits as  expected~\cite{Myers:1986un, Emparan:2008eg}. Our findings suggest that we should classify MP black holes in three series depending on whether the value\footnote{where $n$ denotes the number of angular momenta and $d$ the number of dimensions.}  of $2n - d + 3$ is 0, 1 or 2. For black holes with $2n < d-3$ the Ruppeiner curvature diverges but they have no extremal limits. This is comparable to the recent finding in Ref.~\cite{maria} where the authors are able to establish the minimum temperature surface on which the membrane phase of ultraspinning MP black holes takes place.  In order to allow ultraspinning, \ie infinitely large spin at least one of the possible $\lfloor\frac{d-1}{2}\rfloor$ spins must be exactly $0$. This follows from the general temperature formula (\ref{temperature}) which always allows $T=0$ unless some $J_i$ are zero, see also example of $d = 7$ with three independent spins in Eq.~(\ref{threespins}).

\section{Final remark and conjecture}

We have derived explicit thermodynamic functions of Myers-Perry black holes namely the mass formula in any dimension for the MP black holes with an arbitrary number of angular momentum. We have also derived the Bekenstein-Hawking temperature and the specific heat of the MP black hole in general. It is readily seen that thermodynamic metrics vary with the number of dimensions, $d$ and the number of angular momenta, $n$. We establish  extremal limits of black holes for the MP black holes with arbitrary $n$ in any dimension, in other words we are able to establish the generalized Kerr bound of multiply spinning Kerr black holes in higher dimensions.  We study thermodynamic geometries of the Myers-Perry black holes with arbitrary angular momenta in various dimensions and present the outcomes by drawing thermodynamic cone diagrams which capture the extremal limits of the black holes. Thermodynamic state space can be geometrically represented as a wedge embedded in Minkowski space.  The opening angle of such the wedge is uniquely determined by the number of spacetime dimensions and the number of angular momenta.   We believe that these results will be useful for the purpose of studying higher dimensional black holes. 

We conjecture that the membrane phase ultraspinning MP black holes  is reached at the minimum temperature in the case $2n < d-3$  which is where the Ruppeiner curvature diverges.

\subsection*{Acknowledgments}

Narit Pidokrajt acknowledges the KoF group, Fysikum, Stockholms Universitet for the kind of hospitality. We thank Ingemar Bengtsson for his enlightening and many useful comments.  NP would like to thank Roberto Emparan for enlightening discussions on MP black holes with equal spins while he was a visitor in Barcelona. 


\section*{Appendix A}

We extend our discussion on the opening angles of the Weinhold metrics here with two subcases:

\subsubsection*{The subcase $2n=d-2$}

This applies to even dimensions, 4 with 1 spin, 6 with 2 spins etc. Here
the transformation from $u$ to $\sigma$ is

\begin{equation}
\sigma = \frac{1}{2} \left( \arcsinh\frac{2\ u}{\sqrt{d-3}}
         +\sqrt{d-4}\arctan\frac{2\sqrt{d-4}\ u}{\sqrt{d-3+4u^2}} \right)
\end{equation}

\noindent Here we have $u_{extr} = \frac{\sqrt{d-3}}{2}$ which gives

\begin{equation}
\sigma_{extr} = \frac{1}{2} \left( \arcsinh 1
         +\sqrt{d-4}\arctan\frac{\sqrt{d-4}}{\sqrt{2}} \right)
\end{equation}

\subsubsection*{The subcase $2n=d-1$}

This applies to dimension, 5 with 2 spin, dimension 7 with 3 spins etc. Here
the transformation is

\begin{equation}
\sigma = \frac{\sqrt{d-1}}{\sqrt{2}\sqrt{d-2}}
         \left( \arcsinh\frac{2\sqrt{2}\ u}{\sqrt{d-3}}
               +\frac{\sqrt{d-5}}{\sqrt{2}}
                \arctan\frac{2\sqrt{d-5}\ u}{\sqrt{d-3+8u^2}} \right)
\end{equation}
 
\noindent In this case $u_{extr} = \frac{\sqrt{d-3}}{2\sqrt{2}}$ so we obtain

\begin{equation}
\sigma_{extr} = \frac{\sqrt{d-1}}{\sqrt{2}\sqrt{d-2}} \left( \arcsinh 1
         +\frac{\sqrt{d-5}}{\sqrt{2}} \arctan\frac{\sqrt{d-5}}{2} \right)
\end{equation}

\section*{Appendix B}

In the table below we present extremal limits in various coordinates. 

\begin{center}
\begin{tabular}{|r|r|l|l|r@{ }l|}
\hline
$d$ & $n$ & Extr limits & Extr limits & \multicolumn{2}{c|}{Extr limits} \\
\hline
 4 & 1 & $\frac{S}{J} \geqslant 2$ & $\frac{J}{M^2} \leqslant 1$ &
         $2$ & $\leqslant \frac{S}{M^2} \leqslant 4$ \\[.1cm]
 5 & 1 & $\frac{S}{J} \geqslant 0$ & $\frac{J^2}{M^3} \leqslant \frac{2^4}{3^3}$ &
         $0$ & $\leqslant \frac{S^2}{M^3} \leqslant \frac{2^6}{3^3}$ \\[.1cm]
 5 & 2 & $\frac{S}{J} \geqslant 2$ & $\frac{J^2}{M^3} \leqslant \frac{2^2}{3^3}$ &
         $\frac{2^4}{3^3}$ & $\leqslant \frac{S^2}{M^3} \leqslant \frac{2^6}{3^3}$ \\[.1cm]
 6 & 2 & $\frac{S}{J} \geqslant \frac{2}{\sqrt{3}}$ &
         $\frac{J^3}{M^4} \leqslant \frac{3^\frac{3}{2}}{2^7}$ &
         $\frac{1}{2^4}$ & $\leqslant \frac{S^3}{M^4} \leqslant 1$ \\[.1cm]
 7 & 2 & $\frac{S}{J} \geqslant 0$ & $\frac{J^4}{M^5} \leqslant \frac{2^6}{5^5}$ &
         $0$ & $\leqslant \frac{S^4}{M^5} \leqslant \frac{2^{10}}{5^5}$ \\[.1cm]
 7 & 3 & $\frac{S}{J} \geqslant \sqrt{2}$ &
         $\frac{J^4}{M^5} \leqslant \frac{2^8}{3^3 5^5}$ &
         $\frac{2^{10}}{3^3 5^5}$ & $\leqslant \frac{S^4}{M^5}
           \leqslant \frac{2^{10}}{5^5}$ \\[.1cm]
 8 & 3 & $\frac{S}{J} \geqslant \frac{2}{\sqrt{5}}$ &
         $\frac{J^5}{M^6} \leqslant \frac{5^\frac{5}{2}}{2^2 3^9}$ &
         $ \frac{2^3}{3^9}$ & $\leqslant \frac{S^5}{M^6} \leqslant \frac{2^6}{3^6}$ \\[.1cm]
 9 & 3 & $\frac{S}{J} \geqslant 0$ & $\frac{J^6}{M^7} \leqslant \frac{2^8}{7^7}$ &
         $0$ & $\leqslant \frac{S^6}{M^7} \leqslant \frac{2^{14}}{7^7}$ \\[.1cm]
 9 & 4 & $\frac{S}{J} \geqslant \frac{2}{\sqrt{3}}$ &
         $\frac{J^6}{M^7} \leqslant \frac{3^3}{7^7}$ &
         $\frac{2^6}{7^7}$ & $\leqslant \frac{S^6}{M^7} \leqslant \frac{2^{14}}{7^7}$ \\[.1cm]
10 & 4 & $\frac{S}{J} \geqslant \frac{2}{\sqrt{7}}$ &
         $\frac{J^7}{M^8} \leqslant \frac{7^\frac{7}{2}}{2^{27}}$ &
         $ \frac{1}{2^{20}}$ & $\leqslant \frac{S^7}{M^8} \leqslant \frac{1}{2^8}$ \\[.1cm]
11 & 4 & $\frac{S}{J} \geqslant 0$ & $\frac{J^8}{M^9} \leqslant \frac{2^{10}}{3^{18}}$ &
         $0$ & $\leqslant \frac{S^8}{M^9} \leqslant \frac{2^{18}}{3^{18}}$ \\[.1cm]
11 & 5 & $\frac{S}{J} \geqslant 1$ &
         $\frac{J^8}{M^9} \leqslant \frac{2^{18}}{3^{18} 5^5}$ &
         $\frac{2^{18}}{3^{18} 5^5}$ & $\leqslant \frac{S^8}{M^9}
           \leqslant \frac{2^{18}}{3^{18}}$ \\[5pt]
\hline
\end{tabular}
\end{center}

\end{document}